\documentclass[12pt]{article}

\usepackage{amssymb}

\def\vok{\mathrel{\rlap{\lower0pt\hbox{\hskip1pt{$k$}}}
    \raise6pt\hbox{$\neg$}}}

\begin{document}

\begin{titlepage}

\begin{flushright}
arXiv:1904.13042
\end{flushright}
\vskip 2.5cm

\begin{center}
{\Large \bf Bounds on Vacuum-Orthogonal Lorentz and\\
CPT Violation from Radiative Corrections}
\end{center}

\vspace{1ex}

\begin{center}
{\large Brett Altschul\footnote{{\tt baltschu@physics.sc.edu}}}

\vspace{5mm}
{\sl Department of Physics and Astronomy} \\
{\sl University of South Carolina} \\
{\sl Columbia, SC 29208} \\
\end{center}

\vspace{2.5ex}

\medskip

\centerline {\bf Abstract}

\bigskip

Certain forms of Lorentz violation in the photon sector are difficult to bound directly,
since they are ``vacuum orthogonal''---meaning they do not change the solutions of the
equations of motion in vacuum. However, these very same terms have a unique tendency to
contribute large radiative corrections to effects in other sectors. Making use of this,
we set bounds on four previously unconstrained $d=5$ photon operators at the
$10^{-25}$--$10^{-31}$ GeV$^{-1}$ levels.

\bigskip

\end{titlepage}

\newpage

Since the 1990s, there has been a great deal of renewed interest in the idea
that some seemingly fundamental symmetries in physics, such as Lorentz and CPT symmetries,
might actually be very weakly violated in nature. Thus far, there is no
compelling experimental evidence for such a conjecture. However, since these
symmetries are perceived as being so basic that they underpin our current theories
of elementary particle physics and of gravitation, it is worth understanding exactly how
precisely the symmetries have been measured. If such putatively fundamental
symmetries were found to be ever so slightly broken, that could change our expectations
about the nature of physical laws at the most basic level. Here, we will introduce a new
method for placing strong bounds on forms of Lorentz violation which have previously been
considered quite difficult to observe.

There are a number of reasons why interest in searches for Lorentz and CPT violation have
picked up quite a bit in the last two decades. There has always been the motivation alluded to
above---the notion that any principle that seems so fundamental ought to be studied and understood
as precisely as possible. However, interest in broken Lorentz and CPT symmetries expanded
a great deal as it came to be realized that many of the conceptual frameworks that have been proposed
as ways of describing quantum gravity seem to allow for Lorentz violation, at least
in certain parameter regimes~\cite{ref-kost18,ref-gambini,ref-alfaro,ref-kost20,ref-horava,ref-moffat1}.
Moreover, it was also realized that older tests of Lorentz
and CPT symmetry had not really done a good job of constraining the full parameter space
of Lorentz and CPT violation. With the development of a systematic effective field theory (EFT) to
describe the possible forms that Lorentz violation may take in the interactions of standard model
fields, it became possible to make rigorous comparisons between the results of different types
of Lorentz tests and to design new experiments to explore regions of the parameter space that have
not previously been well constrained.

This EFT is known as the standard model extension (SME)~\cite{ref-kost1,ref-kost2}. Its action
is built using standard model fields, constructed subject to the same requirements as regular
standard model operators, except that they need not be invariant under rotations or Lorentz boosts.
Since it is not possible to have a CPT-violating quantum field theory with a well-defined $S$-matrix
without there also being Lorentz violation~\cite{ref-greenberg}, the SME suffices as the most general
EFT for describing CPT violation as well as Lorentz violation. The SME, in its most general form,
has an infinite tower of operators (of progressively increasing dimension) in its action, but when
the additional requirement of renormalizability is imposed on the theory, the result is the minimal SME
(mSME), which has a finite number of physically observable parameters. In many cases, it makes the
most sense to parameterize the results of experimental Lorentz tests as bounds on linear combinations
of mSME coupling parameters. However, there has also been growing interest in the higher-dimensional
Lorentz-violating operators that are not part of the mSME---especially in the electromagnetic
sector~\cite{ref-kost23}, where precise measurements of the polarization of light coming from cosmologically
distant sources can often been used to place extremely stringent bounds. Depending on what assumptions
are made, many of the same experiments that have been used to bound mSME parameters could also be
interpreted as giving bounds on the coefficients of the non-minimal theory.
A summary of the current bounds on many different SME coefficients may be found in~\cite{ref-tables}.

However, there are some types of higher-dimensional photon Lorentz violation that cannot be bounded using
birefringence measurements. In fact, some parameters are quite difficult to constrain
by any technique. We shall provide new bounds on a class of non-minimal SME operators for photons---operators
which are otherwise quite difficult to study. This will be accomplished through an examination of the
quantum corrections that such operators can contribute to in other sectors of the SME that are easier
to study.

The general form of a SME operator can be constructed as a product of standard model fields and their
derivatives. However, unlike in the standard model, this product may have free Lorentz indices. These indices
are then contracted
with a constant tensor, which represents a preferred background in spacetime. The coefficients that
make up this tensor are the quantities that experiments can be used to place bounds on. If Lorentz symmetry
is broken spontaneously (as might be the case in a number of proposed quantum gravity frameworks), then the
coefficient tensors are related to the vacuum expectation values of dynamical fields that possess tensor indices.

With increasing mass dimension, non-minimal SME operators get more and more derivatives, and
each added derivative typically requires an additional index on the
coefficient tensor, to be contracted with the index on $\partial_{\mu}$.
The proliferation of indices means that the number of possible
Lorentz-violating operators increases with the operator dimension.
Moreover, in most cases, the presence of each additional derivative means an
additional power of the energy in a term's observable effects on relativistic quanta.
However, there are important exceptions; when two
derivatives are contracted to form
$\partial^{\mu}\partial_{\mu}=\partial^{2}$, the resulting term scales as the
invariant mass squared of the quanta, not with the energy. This makes any term in the
electromagnetic sector that includes a $\partial^{2}$ factor what is known
as ``vacuum orthogonal.'' The name comes from the fact that such terms are
unobservable in the vacuum; any vacuum solution of the ordinary Maxwell's
equations (with momentum $p^{2}=0$ in Fourier space), will also be a solution to the theory
with the vacuum orthogonal operator appended to it. This kind of behavior
appears to make the vacuum orthogonal terms hard to constrain. For
example, polarimetric measurements on the radiation from cosmologically
distant sources has been tremendously useful in constraining the
coefficients that lead to photon birefringence~\cite{ref-tables};
the resulting bounds can
be extremely tight, because of the long propagation distances involved,
but the technique cannot be used to constrain the vacuum orthogonal terms.
A completely different approach is needed for the vacuum orthogonal
sector.

Note that discrete symmetries (C, P, and T) cannot be used to distinguish the
vacuum orthogonal operators containing $\partial^{2}$ from rotation-invariant
operators involving $\partial_{0}^{2}$ or $\partial_{j}\partial_{j}$. Thus
far, direct measurements of the vacuum orthogonal terms have typically
utilized matter-filled resonant cavities. The rest frame of
the material breaks the boost invariance and ensures that the photons in the
cavity will have $p^{2}\neq0$.

Yet while the vacuum orthogonal operators have few directly observable tree-level effects, they
(unlike many other terms with $d>4$) are ``unsafe'' with respect to quantum corrections---in the sense that
they can make direct (and large) contributions to the renormalization of
lower-dimensional operators. Higher-dimensional operators with many
Lorentz indices typically cannot make radiative contributions to
$d=3$ and $d=4$ operators, because there are no $d\leq 4$ operators with
the same Lorentz structures. A well known example of this phenomenon
is that when a conventional gauge theory is regulated at short distances
by a lattice, the low-energy behavior is Lorentz invariant,
in spite of the use of a Lorentz-violating regulator. The Lorentz violation due to the
lattice is irrelevant (in the renormalization group sense), because the dominant lattice effects are
characterized by a symmetric four-index tensor, and there are no symmetric four-index
tensor operators in the long-distance theory to ``inherit'' the Lorentz
violation from the short-distance lattice.

However, when a Lorentz-violating operator of dimension $d$ includes a
Lorentz scalar $\partial^{2}$ factor, the operator will have precisely the same
tensor structure as another operator of dimension $d-2$. This means that
the two terms may be intermixed by radiative corrections. We shall
concentrate here on how a $d=5$ generalization of the Lorentz-violating Chern-Simons term in
the photon sector contributes to a $d=3$ $b$-type operator in the charged
fermion sector. However, the phenomenon is fairly general. The radiative
corrections involve virtual photons that are far off shell, with virtual momenta
$p^{2}\sim\Lambda^{2}$, for some ultraviolet cutoff $\Lambda$. So for
every factor of $\partial^{2}$ in the structure of a higher-dimensional operator,
there will be quantum corrections to lower-dimensional analogues that are
enhanced by factors of the large quantity $\Lambda^{2}$.

The SME Lagrange density for a Lorentz-violating generalization of quantum electrodynamics
(QED) with a non-minimal, vacuum-orthogonal, $d=5$ operator takes the form
\begin{equation}
{\cal L}_{5}=-\frac{1}{4}F^{\mu\nu}F_{\mu\nu}+\frac{1}{2}\vok_{5}
^{\mu}\epsilon_{\mu\alpha\beta\gamma}
F^{\alpha\beta}\partial^{2}A^{\gamma}
+\bar{\psi}(i\!\!\not\!\partial-m-e\!\!\not\!\!A)\psi;
\end{equation}
the $\neg$ atop the Lorentz violation coefficient tensor indicates its vacuum orthogonal
nature. The experimental significance of this operator is that it can make radiative
corrections to a readily observable form of Lorentz violation in the fermion sector.
This occurs through the insertion of a $\vok_{5}$ vertex into the
virtual photon propagator in the usual one-loop fermion self-energy diagram.

The contribution this makes to the fermion self-energy is
\begin{equation}
-i\Sigma(p)=-e^{2}\int\frac{d^{4}k}{(2\pi)^{4}}\gamma^{\mu}S(k)\gamma^{\nu}D(p-k)
\left[-i(p-k)^{2}\vok_{5}^{\alpha}(p-k)^{\beta}\epsilon_{\alpha\beta\mu\nu}\right]
D(p-k),
\end{equation}
with $S(k)$ and $D(p-k)$ being the usual fermion and boson propagators. Inserting a Feynman parameter
$x$ and shifting the integration variable to $\ell=k-xp$ gives
\begin{equation}
-i\Sigma(p)=-2e^{2}\epsilon_{\alpha\beta\mu\nu}\int_{0}^{1}dx\, x\int\frac{d^{4}\ell}{(2\pi)^{4}}
\frac{\gamma^{\mu}(\!\not\!\ell+x\!\!\not\!p+m)\gamma^{\nu}\vok_{5}^{\alpha}
[\ell-(1-x)p]^{2}[\ell-(1-x)p]^{\beta}}
{(\ell^{2}-\Delta)^{3}},
\end{equation}
where $\Delta=(1-x)m^{2}-x(1-x)p^{2}$, as usual for the fermion self-energy. For on-shell
fermions, $\Delta=(1-x)^{2}m^{2}$.

The largest radiative
contribution will come from the ${\cal O}(\ell^{4})$ term in the numerator. So the term from
$-i\Sigma(p)$ with the greatest naive degree of divergence is
\begin{equation}
-i\Sigma(p)\sim-2e^{2}\epsilon_{\alpha\beta\mu\nu}\vok_{5}^{\alpha}\int_{0}^{1}dx\, x
\int\frac{d^{4}\ell}{(2\pi)^{4}}
\gamma^{\mu}\gamma_{\rho}\gamma^{\nu}g_{\sigma\tau}\frac{\ell^{\rho}\ell^{\sigma}\ell^{\tau}\ell^{\beta}}
{(\ell^{2}-\Delta)^{3}}
\end{equation}
The $\ell$-integration is quadratically divergent. Cutting off the integration with a regulator at a scale
$\Lambda$ gives
\begin{equation}
\label{eq-Sigma}
-i\Sigma(p)\sim-\frac{3i e^{2}}{64\pi^{2}}\Lambda^{2}\vok_{5}^{\alpha}\gamma_{\alpha}\gamma_{5}.
\end{equation}
This can be found, for example, using dimensional regularization and replacing the divergent $\Gamma$-function
$\Gamma(1-d/2)\rightarrow\Lambda^{2}/m^{2}$; the exact numerical coefficient will depend on the precise meaning of
$\Lambda$ if a different regulator is used, but the result (\ref{eq-Sigma}) will be sufficient for placing conservative,
order-of-magnitude bounds on $\vok_{5}$. This contribution to $\Sigma(p)$ is quite similar that made in a
superficially renormalizable theory with a $d=3$ Chern-Simons term~\cite{ref-kost4}, differing only with the
degree of the divergence.

The net result of this radiative correction is the addition of a term of the form $\bar{\psi}\!\!\not\!b\gamma_{5}\psi$
to the effective Lagrange density for the fermions, where
\begin{equation}
\label{eq-b}
b^{\mu}=-\frac{3e^{2}\Lambda^{2}}{64\pi^{2}}\vok_{5}^{\mu}
\end{equation}
This would have immediate experimental consequences, since the $b$ coefficients give rise to a
number of easily-observed spin-dependent phenomena. Bounds on mSME coefficients such as $b$ are normally
quoted in sun-centered celestial equatorial coordinates $(T,X,Y,Z)$. For
two of the spatial components, $b_{X}$
and $b_{Y}$, the best bounds for a charged fermion species are for protons, made using a He/Xe atomic
magnetometer. The bounds are at the $|b_{X}|,|b_{Y}|\lesssim10^{-32}$ GeV
level~\cite{ref-allmendinger,ref-stadnik}.
The best bounds on $b_{Z}$ (the component along the direction of the Earth's rotation) and
the time component $b_{T}$ come from torsion pendulum measurements using magnetized
samples containing macroscopic numbers of electron spins, with constraints at the
$|b_{Z}|\lesssim10^{-29}$ GeV
and $|b_{T}|\lesssim10^{-26}$ GeV levels~\cite{ref-heckel2,ref-heckel3}.
(These two components of $b$ are harder to measure
because it is not possible to just use the rotation of the Earth to search for their
anisotropic effects.) Unfortunately, none of these bounds are ``clean.'' All the constraints are
on weighted sums of multiple SME coefficients, which are hard to disentangle in nonrelativistic
experiments.
%moreover, the proton limits are also intertangled with the analogous
%coefficients for neutrons.
%So none of these bounds can be expected to have better than order-of-magnitude validity.
However, if $\vok_{5}$ represents the ultimate source of all Lorentz violation in the theory,
then there will be no induced $b$ for a neutral particle like a neutron; and the magnetometer
and torsion pendulum bounds are just on the induced $b$ coefficient for the charged particles---which
is universal by~(\ref{eq-b}).

%Nevertheless,
The extremely tight bounds on linear combinations
containing the $b$ coefficients for charged fermions
should translate into similarly stringent constraints on the $\vok_{5}$ coefficients. However,
to find useful numerical estimates, the issue of the quadratic divergence must be addressed.
The SME, particularly when non-minimal terms are included, must be interpreted as an EFT; it
is valid for calculations up to some scale, but a different ultraviolet completion of the
theory is needed above that scale. Empirically, we know that QED is valid up to approximately
the electroweak scale of $\sim100$ GeV. Nor is there any strong Lorentz violation up to this scale;
this is known from collider searches for Lorentz violation with heavy particles such as the top
quark~\cite{ref-abazov,ref-karpikov}.

We shall therefore take $\Lambda\approx100$ GeV as a lower estimate of the cutoff scale for
the virtual momenta in the fermion self-energy. Based on the existing constraints on fermionic
$b$ coefficients and setting bounds conservatively, at least an order of magnitude looser than given by direct
application of (\ref{eq-b}), to allow for possible cancellations between the radiatively generated
$b$ coefficients and those coefficients that are intrinsic to the fermions, the resulting constraints are
\begin{eqnarray}
\left|\vok_{5}^{X}\right|,\left|\vok_{5}^{Y}\right| & \lesssim & 10^{-31}\,{\rm GeV}^{-1} \\
\left|\vok_{5}^{Z}\right| & \lesssim & 10^{-28}\,{\rm GeV}^{-1} \\
\left|\vok_{5}^{T}\right| & \lesssim & 10^{-25}\,{\rm GeV}^{-1}.
\end{eqnarray}
Not only are these the first bounds placed on the coefficients for vacuum-orthogonal operators with $d=5$,
they are not that dissimilar
to many of the extremely tight bounds on other directly observable $d=5$ operators,
placed using astrophysical birefringence measurements, which cover a $\sim10^{-23}$--$10^{-34}$ GeV$^{-1}$
range~\cite{ref-stecker3,ref-toma,ref-kost30,ref-kislat,ref-friedman}.

This method of placing bounds is specific to the vacuum orthogonal terms. The reason that the
operator parametrized by $\vok_{5}$ is unsafe is that it has the same discrete symmetries and the
same Lorentz structure as the fermion $b$ operators. Other bilinear electromagnetic terms with operator dimension
$d=5$ will have an the same number of derivatives, but without two of the
derivatives being contracted to form a d'Alembertian;
this requires the presence of additional indices on the coefficient tensor, and there are no observable
$d=3$ operators with matching Lorentz structures.

Limits on
%$d>4$
forms of Lorentz violation that are not vacuum orthogonal typically scale as
$E^{-(d-2)}$ or
$E^{-(d-3)}L$, where $E$ is the energy scale of the quanta involved in an experimental measurement
and $L$ is the line of sight in an experiment measuring photons' polarizations or times of flight. For
the vacuum-orthogonal terms in the action, the strength of any bounds based on radiative corrections to
lower-dimensional operators involve the replacement of one factor of $E^{-2}$ by $\Lambda^{-2}$ for
each factor of $\partial^{2}$. Since $\Lambda$ is, in principle, the scale up to which the SME is
valid as a low-energy EFT,
this can indicate significant improvement in the tightness of the bounds, relative to what may be possible
for operators that are not vacuum orthogonal.
This means that radiative corrections are the most natural source for strong bounds on
the vacuum-orthogonal Lorentz- and CPT-violating operators.

The radiative mixing of $\vok_{5}$ and $b$ also works the other direction~\cite{ref-jackiw1,ref-ferrari1};
starting from an action containing a fermion $b$ term, quantum corrections will generate an infinite
series of vacuum-orthogonal photon terms.
Including all the terms at
${\cal O}(b)$, the effective Lagrange density due to $b$ insertions in
the one-loop photon self-energy is
\begin{equation}
\label{eq-DeltaL}
\Delta{\cal L}_{AF}=\frac{e^{2}}{4\pi^{2}}b_{\lambda}\epsilon^{\lambda\mu\nu\rho}
\left\{\left[\frac{\sin^{-1}(\sqrt{-\partial^{2}}/2m)}{(\sqrt{-\partial^{2}}/2m)\sqrt{1+\partial^{2}/4m^{2}}}-1
\right]A_{\mu}\right\}\left(\partial_{\nu}A_{\rho}\right).
\end{equation}
The transcendental function appearing in (\ref{eq-DeltaL}) can be expanded as a Maclaurin series, so
long as $|\xi|<1$, where $\xi=\sqrt{-\partial^{2}}/2m=\sqrt{p^{2}}/2m$. (At $\xi=1$, there is the obvious branch cut,
corresponding to the threshold for the creation of real fermion-antifermion pairs.) The first few terms of
the series expansion are
\begin{equation}
\frac{\sin^{-1}\xi}{\xi\sqrt{1-\xi^{2}}}-1=\frac{2\xi^{2}}{3}+\frac{8\xi^{4}}{15}+\frac{16\xi^{6}}{35}+
\frac{128\xi^{8}}{315}+
%\frac{256\xi^{10}}{693}+\frac{1024\xi^{12}}{3003}+
\cdots.
\end{equation}
(The terms in this expansion suggest, but never quite achieve, a simple pattern.)

The subtraction of the ${\cal O}(\xi^{0})$
term corresponds to having the $d=3$ Chern-Simons term vanish. The radiative
corrections to this term were quite controversial at one point. In a pure Abelian gauge theory,
not embedded in a larger theory that includes gravitation, and with explicit Lorentz
violation in the form of a tree-level $b$ term, it turns out that the $d=3$ term is finite, yet also of
undetermined magnitude. The ambiguity is related to the fact that the $d=3$ term with just one derivative
is not, on its own, gauge invariant. (The integrated action is, however, gauge invariant, which is enough to
ensure that the equations of motion are also invariant.) The ambiguity does not extend to the non-minimal terms,
since they are fully gauge invariant, depending only on derivatives of $A$ rather than on $A$ itself.

Moreover, if the Abelian gauge theory is part of a larger theory that includes general relativity (or
a more general metric theory of Riemannian gravity), matters are subtly different. Explicit breaking of
Lorentz invariance by a $b$ term is not consistent
with the metrical structure of spacetime~\cite{ref-kost12}.
Unless $b$ is actually derived from the vacuum expectation value of
a separate axial vector field (endowed with its own nontrivial dynamics), the Bianchi identities that are
required for the geometric interpretation of gravity cannot be satisfied. (There may be more general geometric
theories of gravitation, perhaps utilizing Finsler geometry, that avoid this problem. However,
these potential theories are, at present, too poorly developed to provide a framework for studying loop corrections
involving quantum fields on these kinds of backgrounds.) The modified photon self-energy must be transverse
to both of the external momenta, which are potentially different, since $b$, being a dynamical quantity, can
carry momentum itself~\cite{ref-altschul37}.
The presence of two independent transversality conditions forces the self-energy to
be at least quadratic in momentum, which rules out the $d=3$ operator, but not the $d>4$ ones~\cite{ref-coleman}.

Unfortunately, the radiative corrections to $\vok_{5}$ do not seem to be useful for setting any additional bounds.
The coefficients $\vok_{5}$ of the $d=5$, vacuum-orthogonal form of Lorentz violation are extremely difficult
to constrain directly. What we have shown, however, is that $\vok_{5}$ and similar terms are, on the other hand, uniquely
susceptible to being constrained using radiative corrections. Existing bounds on the $b$ parameters for charged
fermions can be interpreted as limits on the components of $\vok_{5}$ at the $10^{-25}$--$10^{-31}$ GeV$^{-1}$ levels,
not so different from the bounds on other $d=5$ operators which are not vacuum orthogonal.

\end{document}